\documentclass[journal,12pt,onecolumn,draftclsnofoot,]{IEEEtran}
\IEEEoverridecommandlockouts
\usepackage{cite}
\usepackage{comment}
\usepackage{amsmath,amssymb,amsfonts}
\usepackage{graphicx}
\usepackage{textcomp}
\usepackage{xcolor}
\usepackage{tabularx}
\usepackage{multirow}
\usepackage{color}
\usepackage{colortbl}
\usepackage{tikz}
\usepackage{float}
\usepackage[noend]{algpseudocode}
\usepackage[linesnumbered,ruled,vlined]{algorithm2e}
\usepackage{amsmath,amsfonts}

\SetCommentSty{mycommfont}
\SetKwInput{KwInput}{Input}                
\SetKwInput{KwOutput}{Output} 
\def\BibTeX{{\rm B\kern-.05em{\sc i\kern-.025em b}\kern-.08em
    T\kern-.1667em\lower.7ex\hbox{E}\kern-.125emX}}
\usepackage{url}

\begin{document}

\title{Attack Prediction using Hidden Markov Model}

\author{\IEEEauthorblockN{Shuvalaxmi Dass, Prerit Datta, Akbar Siami Namin \\}
\IEEEauthorblockA{\textit{Department of Computer Science} \\
\textit{Texas Tech University}\\
Lubbock, Texas, USA \\
shuva93.dass@ttu.edu; prerit.datta@ttu.edu; akbar.namin@ttu.edu}
}

\maketitle

\begin{abstract}
It is important to predict any adversarial attacks and their types to enable effective defense systems. Often it is hard to label such activities as malicious ones without adequate analytical reasoning.  
We propose the use of Hidden Markov Model (HMM) to predict the family of related attacks. Our proposed model is based on the observations often agglomerated in the form of log files and from the target or the victim's perspective. We have built an HMM-based prediction model and implemented our proposed approach using Viterbi algorithm, which generates a sequence of states corresponding to stages of a particular attack. As a proof of concept and also to demonstrate the performance of the model, we have conducted a case study on predicting a family of attacks called \textit{Action Spoofing.}
\end{abstract}


\section{Introduction}
\label{sec:introduction}

Hidden Markov Model (HMM) is a probabilistic model that can be used to predict a set of ``\textit{hidden}'' states based on certain observations. The foundation of Hidden Markov Model (HMM) was initially introduced 
with its applications primarily used in speech and signal processing. Nowadays, HMM finds its applications across several domains such as natural language processing and machine-learning due to the simplicity of adapting the model to predict unknown (i.e., hidden) sequence of states. The prediction is based on the features or observations emitted from each state. HMM are very useful in cyber security domain as cyber-attacks are often conducted in several phases or steps where these steps may not always be conspicuous as attackers often try to mask their activities. However, HMM can help in identifying patterns in the data such as network trace spread across time and can evidently help in determining attacks \cite{Holgado2020}. HMM also find its applications in cloud computing to detect multi-step attacks using the cloud \cite{Chen2012}. It also can be used to detect attacks that target misconfigured software applications \cite{be3899e815be45888cfc7ee529351b22} \cite{9202661}.

Cyber-attacks are often conducted in several phases or steps, which may not always be conspicuous, as the attackers often try to mask their activities by following a stealth approach. HMM is an effective technique in unmasking the attacker's intentions by correlating the temporal observations spread across network and activities' log files. HMMs have been widely used in combination with intrusion detection systems (IDS) as an early-warning system for attack detection or for detecting anomalies in the network. 

In this paper, we propose to adopt Hidden Markov Model to predict attacks. The introduced model consists of hidden states and observations related to the family of attacks under consideration. To demonstrate the formulation and modeling of the HMM-based attack prediction, the results of a case study will be presented with its focus on action spoofing. 

The case study considers user's activities that are directly observable (e.g., user opens an app) and the attacker's malicious activities running in the background which are hidden to the user. Moreover, we are dealing with attacks belonging to the same family instead of different families. Defining a HMM model for attacks falling under varied families can get infeasible due to a very large search space which is computationally infeasible. The motive to consider an attack family is to simplify the problem by grouping attacks with similar properties and characteristics. As a result, a defense strategy employed to confront a certain type of attack or mitigate the risk of such attack can be adapted for similar attacks classified in the same family. The observations are chosen from the user or the user's perspective (i.e., victim) as we want to assess if the observations can be successfully utilized in determining the corresponding attack family. The key contributions of this paper are:

\begin{enumerate}
    \item We introduce an attack prediction model using Hidden Markov Model in which attacks are grouped based on their characteristics and thus, prediction is performed on the attacks belonging to the same family. 
    \item We demonstrate the process of the introduced model in predicting attacks through a case study where observations are  derived from analysis of log files in the context of predicting a family of attacks belonging to action spoofing. 
\end{enumerate}

This paper is structured as follows: The literature of this line of research is reviewed in Section \ref{sec:relatedwork}. Section \ref{sec:background} briefly presents the technical background of Hidden Markov Model (HMM). The problem of attack prediction using HMM is presented in Section \ref{sec:problem}. Section \ref{sec:HMMforAttack} presents the  methodology  of  how  attacks  are  modeled  into  the  hidden  markov  model. The use of Viterbi algorithm to determine the type of attack based on observations is presented in Section \ref{sec:algorithm}. The results of a simulation are presented in Section \ref{sec:simulation}. The practical implications of the approach are discussed in Section \ref{sec:practicalImplication}. The conclusion of the paper along with some future work are provided in Section \ref{sec:conclusion}.

\section{Related Work}
\label{sec:relatedwork}



Zan et al. \cite{Zan2009} present a framework to detect cyber attacker's intentions based on HMM called HMM-AIP (Attack Intention Prediction). The authors train several HMMs to predict seven types of attack intentions based on alerts sequence received from the IDS. The authors consider six types of attacks scenarios 1) privilege escalation, 2) worm, 3) botnet, 4) phishing, 5) Web attack and 6) DDoS to train the HMM with the goal of detecting attacker's intentions based on the sequence of attack alerts from the IDS. The authors evaluate their framework using the DARPA 2000 intrusion dataset to map the alerts to attack intentions. 

HMM has also been used to predict the attack sequences in the cloud computing environment. The detection of attacks in the cloud is vital as the attackers can use the cloud computing resources to launch larger attacks. Chen et al. \cite{Chen2012} propose a system based on HMM to detect attacks in the cloud through analysis of log files. The assumption is that the malicious activities by the attacker often leave some traces at each attack step, which can be analyzed from the logs files. 

Kholidy et al. \cite{Kholidy2014} presents a risk-centric approach to predict multi-stage attack in the cloud. The authors use an HMM model to predict the attack sequence from the sequence of alerts generated by the IDS. The HMM model is trained to recognize attack signatures based on combinations of alerts relevant to the attack. 
The authors conclude that the HMM model combined with IDS was able to generate early-warning alerts for severe attacks by 39 minutes before the attack phase.

Holgado et al. \cite{Holgado2020} propose a method based on HMM to predict multi-stage attacks utilizing the alert from the IDS. The authors use a clustering method that tags alerts with certain keywords. The textual description of the alerts is used to extract significant keywords utilizing the CVE repository. Once tagged, these alerts are used as observations for training the HMM model to determine the multistep attack sequence. 

Besides HMM, there are some other techniques for attack predictions such as Bayesian networks \cite{Ahmet2017}, decision trees \cite{Sanjana2020}, hidden petri nets \cite{Yu2007}, data-intensive Artificial Neural Networks (ANN) and deep learning \cite{dong2020}. Fuzzy and game theoretic approaches for attack prediction are useful for the IDS systems to take a defensive action in response to when an attack is detected \cite{Holgado2020}. Whereas, hidden petri colored nets \cite{Yu2007} require pre and post-conditions of an attack in order to predict it.

Unlike hidden Petri nets, HMM allows prediction of attacks based on observations from each state and does not require a large set of conditions to build the model \cite{Holgado2020}. In addition to performance benefits of HMM over decision trees \cite{Sanjana2020}, HMM does not need a lot of data compared to ANN and deep learning models \cite{dong2020}.
The work presented in this paper uses ``{\it textual description}'' of security attacks to formulate hidden states and observations which is different than existing techniques such as measuring the correlation of security events to predict attacks. Furthermore, our proposed model predicts related but different attack types and can be further adapted to any multi-step attacks.

\section{A Brief Background of HMM}
\label{sec:background}


\subsection{Hidden Markov Model}
Hidden Markov Model (HMM) is a probabilistic model and it is kind of an extension to Markov chains. A Markov Chain is a stochastic model that describes transition between events or states based on some probability values. Unlike Markov chains, the states in a HMM are not directly observable and thus are ``hidden.'' The states can be estimated based on the emissions or observations at each state.

Like Markov chains, HMM embodies two important assumptions about the model. The first property also known as ``\textit{markov assumption}'' states that given a sequence of states \(X_0, X_1,... X_n \), the future state $X_{n+1}$ is only influenced by the current state $X_n$: 

\vspace{-0.15in}
\begin{align*}
Pr(X_{n+1} = j | X_n = i, X_{n-1} =a,\dots, X_0 = z) &= \\Pr(X_{n+1} = j | X_n = i)
\end{align*}

Additionally, the probability of transition from state i to state j remains the same throughout the sequence. This property is also know as ``\textit{time invariance}'' or ``\textit{time stationary}'' property (i.e., $Pr(X_{n+1} = j| X_n = i)$ remains same $\forall n = 0,1,2\dots$ ).

An HMM consists of the following components \cite{baum1966statistical}:

\begin{itemize}
    \item[--] A set of finite \textbf{states} $X= \{X_1,X_2,X_3,\dots, X_n \}$. These states are not directly observable and are \textit{hidden}. Instead, each state emits or outputs a unique symbol which can be used to estimate the state.
    \item[--] A set of \textbf{observations} or \textbf{emissions} $Y = \{Y_1,Y_2,\dots,Y_n\}$, where $Y_i$ are distinct symbols emitted by each state $X_i$ at different time intervals. 
    \item[--] A \textbf{state-transition} matrix $A = a_{11},a_{12},\dots a_{ij}$ that describes the probability of transitioning from state $i$ to state $j$ such that, 
        \begin{math} 
            \sum_{j=1}^{n} a_{ij} =  1 \quad 1\leq i,j \leq n. 
        \end{math}
    \item[--] An \textbf{observation probability} matrix $B = b_i(Y_t)$, where $b_i$ is the probability of observation emitted $Y_t$ by state $i$.
    \item[--] An \textbf{initial probability distribution}  $\pi = \pi_1,\pi_2,\dots\pi_n$ where $\pi_i$ is the probability that the model starts at state $i$. It should be noted that 
    \begin{math} 
            \sum_{i=1}^{n} \pi_i =  1.
        \end{math}
\end{itemize}
Together, these components of the HMM can represented by the notation $\lambda (A,B,\pi)$.

\subsection{Fundamental Problems Modeled by HMMs}

Rabiner \cite{hmmtutorial} describes three fundamental problems that can be addressed using hidden Markov models:

\begin{enumerate}
  \item \textbf{Evaluation Problem.} Given the observations $Y = \{Y_1,Y_2,\dots,Y_t\}$ and the HMM $\lambda (A,B,\pi)$, how can we deduce $P(Y|\lambda)$ efficiently? More specifically, given an HMM  and a sequence of observations how likely is it that the given HMM  results in those observations?
  \item \textbf{Decoding Problem.} Given the observations $Y = \{Y_1,Y_2,\dots,Y_t\}$ and the HMM $\lambda (A,B,\pi)$, how to find a state sequence $X= \{X_1,X_2,X_3,\dots, X_n \}$ that best corresponds to the observations. The key idea is to find the correct or most probable sequences of hidden states of the HMM based on sequence of observations values.
  \item \textbf{Optimization Problem.} Given an HMM $\lambda$ with parameters $A,B$ and $\pi$, how to optimize the parameters to maximize $P(Y|\lambda)$? This process is also known as ``\textit{training}'' as the observed sequences are used to adjust (fine-tune) the values of the model's parameters
\end{enumerate}

The first problem can be solved by an algorithm known as \textit{Forward-Backward Procedure}. The second problem can be resolved using \textit{Viterbi algorithm} and finally, the solution to the third problem can be achieved by \textit{Baum-Welch} or \textit{EM (Expected-Maximization)} methods. In this paper, we apply the HMM to our application domain to predict the sequences of attacker's state in the attack execution flow based on the observations from the users actions (i.e., an instance of the second problem). The attack's hidden states are then used to predict the attack type. We discuss how we formulated Viterbi algorithm 
in the next section.

\section{PROBLEM FORMULATION}
\label{sec:problem}

It is often important to find the states that produces a given observation sequence. For example, given a patient's symptoms monitored over a period of time, a doctor may wish to find what disease resulted in the observed set of symptoms. An HMM can be used for such applications. The Viterbi algorithm tries to determine the best sequence of hidden states that resulted in the observed sequences. Viterbi algorithm is an application of dynamic programming and it uses recursion to determine the hidden state sequence given the observations. It uses a graph structure called ``\textit{trellis}'' consisting of various paths connecting the start state to other hidden states at different time intervals. The Viterbi algorithm then estimates the likelihood of following a path given the input sequence of observation at each time step. The following subsection describes formulation process used in our domain.

\subsection{Viterbi Algorithm}
Given a particular sequence of observations $Y = \{Y_1,Y_2,\dots,Y_t\}$ and the HMM $\lambda (A,B,\pi)$, the Viterbi algorithm aims to find an optimal hidden state sequence $P(X|Y,\lambda)$ associated with the given observation sequence. In the context of this paper, $Y_i$ is the instance or event that is observed or seen by the user/victim at a time instant \textit{t} with a malicious action running in the background unknown to the user (i.e. hidden state $X_i$). The tasks of logging into the system and receiving email links are examples of such events. Once we define the model parameters ($A, B,$, $\pi$), we then define \cite{hmmtutorial}:

\vspace{-0.12in}
\begin{equation}
    \alpha_t (i) = \max_{X_1,X_2,\dots,X_{t-1}} P[X_1 X_2 \dots X_t = i, Y_1 Y_2\dots Y_t | \lambda]\label{eq:1}
\end{equation}

where $\alpha_t(i)$ represents the Viterbi variable which corresponds to the highest probability of a single path given the observation sequence till time interval $t$. More specifically, Equation \ref{eq:1} attempts to find that hidden state sequence: $X_1, X_2, \dots X_t = i$ whose probability of emitting the given observation sequence: $ Y_1 Y_2\dots Y_t$  is maximum. Here, $X_i$ represents the hidden state at time interval t. In other words, the algorithm tries to find a sequence of states (events) that are most likely to produce the input observation sequence, given the HMM model $\lambda$. 

In our problem domain, $\alpha_t$ would be high for that attack path (i.e., hidden state sequence), where the given observation sequence closely resembles the outcome of execution flow (events) in that attack path.  
For time $t+1$:

 \vspace{-0.12in}
\begin{equation}
    \psi_{t+1}(j) = \alpha_{t+1}(j) = [\max_i \alpha_t(i)a_{ij}] . b_j(Y_{t+1})
\end{equation}


where:
\begin{itemize}
    \item [--]  $\psi_{t+1}(j)$ is used to keep track of the max value for Equation \eqref{eq:1}, 
    \item [--]  $b_j$ is the probability of observation $Y$ emitted  by hidden state $j$ at time step $t+1$,
    \item [--]  and $a_{ij}$ is the probability of transitioning from state $i$ to $j$. 
\end{itemize}

To derive the state sequence, we need to keep a track of the values that corresponds to the highest probability of a single path for each $t$ and state $j$. The implementation of Viterbi algorithm can then be specified through four steps \cite{hmmtutorial}:

\begin{enumerate}
    \item {\it Initialization}: 
        \begin{subequations}
        \begin{align} 
        \alpha_1(i) &= \pi_i b_i(Y_1), 1 \leq i \leq n \\
         \psi_1(i) &= 0
         \end{align}
        \end{subequations}
        We initialize the Viterbi variable by multiplying the initial probability of state $i$ (i.e., $\pi_i $), with the emission probability of state $i$ to $Y$ at time  $t = 1$ (i.e., $b_i(Y_1)$).
    \item {\it Recursion}: 
        \begin{subequations}
        \begin{align} 
        \begin{split}
            \alpha_t(j) = \max_{1 \leq i \leq n} [\alpha_{t-1}(i)a_{ij}]b_j(Y_t), \quad 2 \leq t \leq T \\
            1 \leq j \leq n
        \end{split}
         \end{align}
        Here, the maximum value among the multiplication results is calculated. A single multiplication comprises of: 1) previous Viterbi variable of state $i$: $\alpha_{t-1}(i)$, 2) the transition probability from state $i$ to $j$: $a_{ij}$, times 3) the emission probability from state $j$ to observation $Y$: $b_j(Y_t)$  and assign that to the new Viterbi variable.
        \begin{align} 
            \begin{split}
            \psi_t(j) = \arg\ max_{1 \leq i \leq n} [\alpha_{t-1}(i)a_{ij}], \quad 2 \leq t \leq T \\
            1 \leq j \leq n
            \end{split}
         \end{align}
          \end{subequations}
        The $\psi$ array is used to save those state entries that maximized the Viterbi variable. 
    \item {\it Termination}: 
        \begin{subequations}
        \begin{align} 
            P^* &= \max_{1 \leq i \leq n} [\alpha_T(i)]\\
            X^*_T &= \arg\ max_{1 \leq i \leq n} [\alpha_T(i)]
         \end{align}
         \end{subequations}
          It represents the probability of the best entire state sequence. In the context of this paper, $P^*$ gives the best path probability (attack path) while $X^*_T$ describes the actual hidden states corresponding to the best attack path. 
    \item {\it Path Backtracking}: 
        \begin{align} 
            X_t^* = \psi_{t+1}(X_{t+1}^*), \quad t = T - 1, T - 2, \dots,1
        \end{align}
         This step finds the hidden state sequence by  tracing back through the $\psi$ arrays.
\end{enumerate}


\section{Attack Prediction Through HMM}
\label{sec:HMMforAttack}

In our methodology, given a sequence of observations as input, our aim is to use Viterbi algorithm to generate a sequence of hidden states which is representative of the \textit{attack flow}. We then identify the attack type whose execution flow resembles the resulted sequence.

\subsection{Attacks Description}
As a proof of concept, we study a family of \textit{Action Spoofing} attacks and thus predict them using HMM. In these types of attacks, an attacker tricks a user into performing certain malicious actions unbeknownst to the user (e.g., clicking a button that secretly downloads a malicious application). We referred to the Common Attack Pattern  Enumeration and Classification (CAPEC) \cite{capec, Chin2011} for the attack descriptions as it is widely used as a reliable repository of attack patterns and their descriptions. The specific attack types that fall under Action Spoofing are: 

\begin{enumerate}
    \item \textit{Clickjacking}: In these type of attacks, the attacker often overlays a transparent user-interface (UI) on the top of the existing UI to mask certain areas that trigger malicious actions when clicked. 
    
    \item \textit{Activity Hijack}: This type of attack is often common to android operating systems. Every app on android needs to register an intent and explain what the app does when certain actions are triggered. An attacker causes a malicious activity to trigger instead of a trusted activity by registering an implicit intent and thereby prompting the user to enter sensitive information. 
    
    \item \textit{Task Impersonation}: A malicious application installed on the user's computer secretly monitors the process list to trigger a malicious application resembling an actual application that the user wants to open. The user is then tricked into interacting with the malicious application and can expose sensitive information if the application requires the user to enter credentials.
    
    \item \textit{Scheme Squatting}: In this type of attack, a previously installed malicious application registers a URL scheme on behalf of a target application that is not installed yet on the user's system. When the target application is installed, the malicious application displays an interface resembling that of the target application, and thus tricking the user to input sensitive information.
    
    \item \textit{Tapjacking}: Similar to clickjacking attack, a previously installed application displays an interface resembling a benign application and the user is tricked into \textit{tapping} a certain area of the malicious application which may further trigger malicious actions or activities.
\end{enumerate}

\subsection{Hidden States}
To represent the hidden states as attack flows, we utilized both the textual \textit{descriptions} and the \textit{pre-requisites} defined for every attack type in the CAPEC website \cite{capec} and encoded them as keywords into discrete hidden states. Table \ref{tab:states_table} shows the complete list of the hidden states for our HMM model. 

\begin{table}[]
\centering
\caption{List of hidden states representing attack flow instances for all attack types in the HMM model.}
\label{tab:states_table}
\resizebox{4.5cm}{!}{
\begin{tabular}{cl}
\hline
\textbf{States} & \multicolumn{1}{c}{\bf Description}         \\ \hline
S1              & Installed malicious s/w      \\ 
S2              & Flash App Overlay            \\ 
S3              & URL scheme registering       \\ 
S4              & Implicit intent interception \\ 
S5              & Query legit task list        \\ 
S6              & Toast Window overlay         \\ 
S7              & Message interception         \\ 
S8              & Launches new task            \\ 
S9              & Mimics trusted UI            \\ 
S10             & User tricked                 \\ 
S11             & Sensitive data obtained      \\ \hline
\end{tabular}}
\vspace*{-0.2in}
\end{table}

As we are dealing with just one family of attack (i.e., \textit{Action Spoofing}), its specific attack types have a lot of overlapping among events (e.g., compromising the victim's data). We accounted for these overlapping instances by representing them as single states in our HMM model, for instance, \texttt{S10, S11} in Table \ref{tab:states_table} are some of the events which were redundant in their occurrences across different attack paths.

\subsection{Observations}
\label{subsec:observations}

The observations in our HMM model are instances or events that are observed or seen by the user as a series of sequential events but with a malicious action running in the background without the knowledge of the user. To ease describing the process, we illustrate an attack scenario to provide the details explaining the motive of the attacker. We then build upon the scenario and generate observations pertaining to different attack types. 

{\it An Illustrative Attack Scenario.} Consider a victim user who logs into their bank account 
through a Bank application on their android device. We assume that a  malicious app has already been installed on the victim's device and is disguised as a genuine application. On opening the malicious app, user is prompted with a message link (e.g., \textit{`Claim free gift!'}), which redirects the user to a malicious website disguised as a genuine one. The intent of the attacker is to trick the user into authorizing a bank transaction by manipulating the user interface such that the user remains oblivious to the malicious activity which results in unauthorized money transfer to the attacker's account.

Depending on the attack type and its family, the mechanism to trick the user varies. However,  the starting and ending points, (e.g., visiting or installing a website or malicious app) (State $S1$) and stealing money (State $S11$), respectively  are the same since all the attack types belong to the same family with same goal of tricking the user . Table \ref{tab:obs_table} shows the list of observations we considered based on what user sees and performs. This list incorporates all the activities carried out by different attack types.

\begin{table}[]
\centering
\caption{List of observations emitted by every state present in HMM.}
\label{tab:obs_table}
\begin{tabular}{cl}
\hline
\textbf{Obs.} & \multicolumn{1}{c}{\bf Description}                       \\ \hline
O1                   & User logged in                               \\ 
O2                   & Receives email link (Malicious)                   \\ 
O3                   & Visits Web pages; clicks on overlay button   \\ 
O4                   & Unauthorized account transaction           \\ 
O5                   & opens malicious app                            \\ 
O6                   & Toast window pops up; taps on button       \\ 
O7                   & Malicious App running in background              \\ 
O8 & Clicks on unregistered URL for uninstalled Bank App\\ 
O9                   & Sign-In look-alike page open on foreground \\ 
O10                  & Enters card details                        \\ 
O11                  & Launches new Bank App task                 \\ 
O12                  & Launches legit charity App                 \\ 
O13                  & Selects Bank App from options              \\ \hline
\end{tabular}
\end{table}

\section{Algorithm}
\label{sec:algorithm}

Algorithm \ref{alg:VA} shows the procedure for implementing Viterbi Algorithm in our problem domain.

\begin{algorithm}[]
\DontPrintSemicolon
  \KwInput{
          1) State Transition Matrix $A = [N \times N]$
           \newline
          2) Observation Matrix $B = [N \times M]$
           \newline
          3) Initial probability distribution $\pi = [1 \times N]$
          \newline
          4) Hidden states $hidden\_states = \{s_1,s_2,\dots, s_n \}$
          \newline
          5) Observation sequence $O= \{o_1,o_2,\dots, o_j \}$ 
           }
  \KwOutput{Attack path Sequence $A= \{s_1,s_2,\dots, s_k \}$ }
  {
    p\_list = []  \Comment{\textit{To store probability values calculated in line 2}}
    \newline
    attack\_path = []
    \newline
    probabilities = [] \Comment{\textit{To store tuples of p\_list as shown in line 3}}
    \newline
    }
   \For {($states$ in $hidden\_states$)}
        {
            p\_list.append($\pi$[states] *    $B$[states][$o_1$])
        }
        probabilities.append($tuple$(p\_list))
        
    \For {($obs$ = $o_2$ to $o_j$)}
        {
          prev\_states = probabilities[-1]
          
         max\_prob\_tuple[obs] = max probabilities for all hidden states using $prev\_states$ probs for $obs$ 
        
        probabilities.append(max\_probs\_tuple[obs])
        }
     \For {($p\_tuple$ in $probabilities$)}
          {
            max\_p =  max prob in $p\_tuple$
            attack\_path.append($hidden\_state$ associated to max\_p)
          }
     \textbf{print} (attack\_path)
     
\caption{Identifying state sequence based on given observation sequence  using Viterbi Algorithm}
\label{alg:VA}
\end{algorithm}
 
The input to the algorithm is a defined HMM model $\lambda (A,B,\pi)$ and a sample observation sequence $O = \{o_1,o_2,\dots, o_j \}$. Here $A$ represents the \textbf{state transition probability (TP)} matrix of size $11 \times 11 (=N)$ as there are 11 hidden states in our model (Table \ref{tab:states_table}), $B$ represents \textbf{emission probabilities (EP)} matrix of size ($11 \times 13)$ as we have $13 (=M)$ distinct observations (Table \ref{tab:obs_table}), and lastly, \textbf{initial probability distribution (IP)} vector $\pi$ of size $ 1 \times 11$ which represents the probability of choosing  particular hidden state as the starting point. The pseudo code described in the algorithm is adapting Python syntax.
The output of the algorithm is an $attack\_path$ sequence which consists of hidden states. This algorithm consists of the following steps:
\begin{enumerate}
    \item \textbf{Lines 1 - 3:} We created a tuple consisting of probabilities for the $1^{st}$ observation $o_1$ in O which is calculated by multiplying $IP$ (i.e., Initial Probability) of each hidden state  by their $EP$ (i.e., Emission Probability) for $o_1$ and then append the tuple to the $probabilities$ list. 
    
    \item \textbf{Lines 4 - 7:} For every $obs$ in O[$o_2, ..., o_j$], we obtain the last tuple of probabilities i.e., $probabilities[-1]$ which gives us probabilities of all previous hidden states  and  assigns it to the previous hidden states list, $prev\_states$. We then identify the maximum from the probabilities of each of the next hidden states based on all previous hidden states probabilities. For instance, for a particular hidden states, $hs$ and observation $obs$, the maximum probability is chosen among $all$ hidden states considered one at a time from 1 to N as follows: 
    
    \[
    max(prev\_state\_probs[1:N]*TP[1:N][hs]*EP[hs][obs])
    \]
    
     Where $TP$ is the Transition Probability. The tuple consisting of maximum probability for a particular $obs$,\\ $max\_prob\_tuple[obs]$, for all $hs$: 1 to N is then added to $probabilities$ list and the entire process keeps repeating until we reach the last $obs$ on $O$.
     
    \item \textbf{Lines 8 - 9:} We then iterate over each tuple, ($p\_tuple$) in $probabilities$ list and identify which hidden state has the max probability among the rest. Here, each tuple corresponds to its respective $obs$ in $O$. The hidden state is then added to the attack path. The process is repeated until we reach the last tuple or $obs$.
    
 \item \textbf{Line 10:} We eventually obtain the best attack path associated to the observation sequence $O$.
     
\end{enumerate}
This way we could obtain the best attack path out of all the possible paths that resulted in the given observation sequence.

 
\section{Simulation and Results}
\label{sec:simulation}
We implemented the HMM and Viterbi algorithm using Python version 3.6.
This section reports the results of a simulation performed to demonstrate the mechanic of the proposed HMM-based model for attack prediction.

\subsection{Data Collection}
We collected our data from CAPEC's website for the {\it action spoofing}\footnote{https://capec.mitre.org/data/definitions/173.html} family of attacks. For each attack, we utilized the attack's description and execution flow to derive the states and observations to create our model. As an example, consider the description of \textit{tapjacking} attack below:

\begin{quotation}
    {\it ``An adversary, through a previously installed malicious application, displays an interface that misleads the user and convinces them to tap on an attacker desired location on the screen. This is often accomplished by overlaying one screen on top of another while giving the appearance of a single interface. There are two main techniques used to accomplish this. The first is to leverage transparent properties that allow taps on the screen to pass through the visible application to an application running in the background. The second is to strategically place a small object (e.g., a button or text field) on top of the visible screen and make it appear to be a part of the underlying application. In both cases, the user is convinced to tap on the screen but does not realize the application that they are interacting with.''}
\end{quotation}

The intention of the action spoofing attack is to trick the user into performing certain malicious actions. The tapjacking attack (a type of action spoofing) in particular does this by using a toast overlay window to trick the user into clicking certain section of the screen, which has an underlying benign app running.

Similar to our illustrative example described in Section \ref{subsec:observations},  the previously installed malicious app sends a prompt ('claim free gift') which is an invisible overlay window on the top of the bank app running in the background. When the user clicks on the claim free gift button, which is exactly aligned over the login button in the banking app, an unauthorized transaction is initiated. Based on the tapjacking and the attack description, we derived the hidden states which are from the attacker's perspective and the corresponding observations which are from the user's perspective, as listed in Table \ref{tab:tapjacking}.

\begin{table}[!ht]
{\footnotesize
\centering
\caption{Tapjacking: states and observations.}
\label{tab:tapjacking}
\begin{tabular}{|l|p{4.5cm}|}
\hline
\multicolumn{1}{|c|}{\textbf{Hidden states}}           & \multicolumn{1}{c|}{\textbf{Obseravations}}                   \\ \hline
S1:Installed malicious software & O1:User logged in                       \\ \hline
S6:Toast window overlay         & O5:opens malicious app                  \\ \hline
S10:User tricked               & O6:Toast window pops up; taps on button \\ \hline
S11:Sensitive data obtained     & O4:Unauthorized account transaction    \\ \hline
\end{tabular}
\vspace{-0.12in}
}
\end{table}



\subsection{Probability Values for HMM Parameters}

 We defined our HMM model by constructing the probability matrices: $A$, $B$ and $\pi$. We estimated the probability values using certain heuristics derived from each attack’s description and execution flow described on the CAPEC website \cite{capec}.

For instance, in scheme squatting attack \cite{capec}, the device/computer should have a pre-installed malicious software (i.e hidden state S1). The malicious software allows the attacker to register for a URL scheme (i.e hidden state S3)  meant for a target application. Hence, state S2 is more likely to occur next from state S1 for that attack path.

\begin{figure}[!t]
 \centering
  \includegraphics[width=\linewidth]{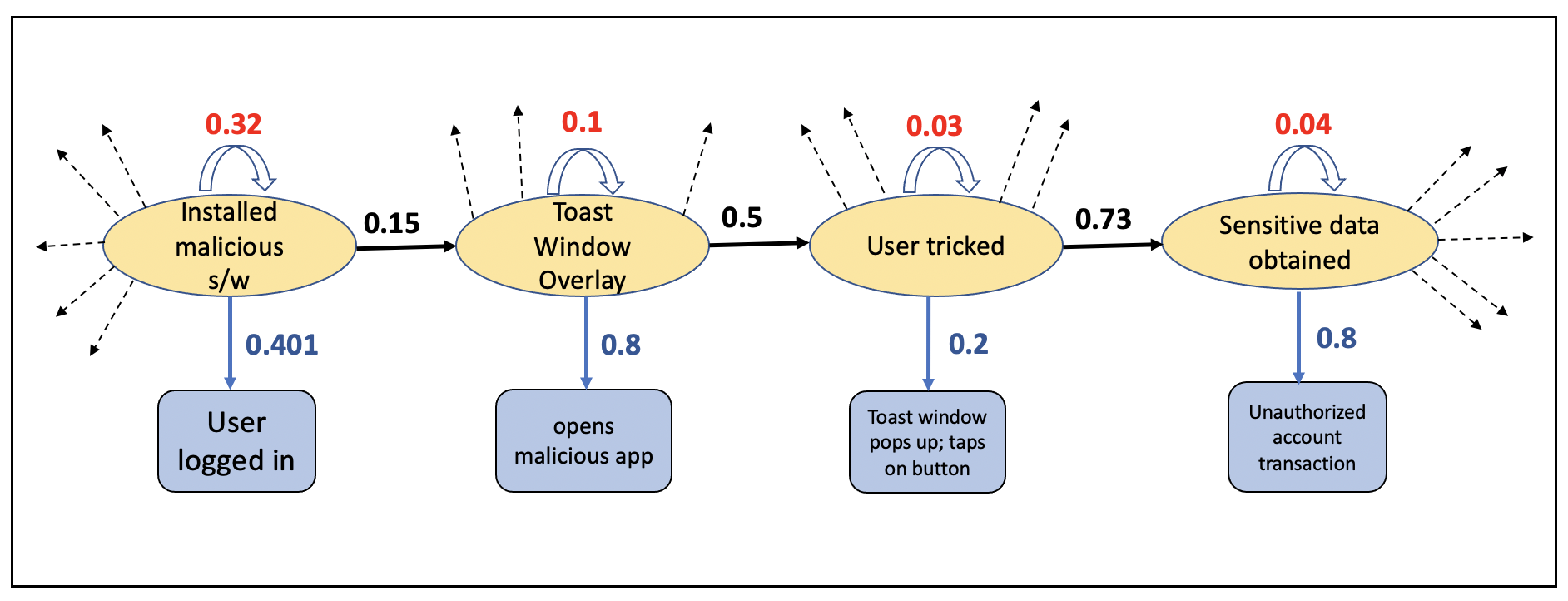}
  \caption{State transition diagram for Tapkjacking with initial, transition and emission probabilities.}
  \label{fig:clickjacking}
  \vspace*{-0.15in}
\end{figure}

Furthermore, to derive the value of $\pi$, it was apparent to give the state \texttt{S1: Installed malicious s/w} the highest starting probability as all attacks are contingent upon a pre-existing malicious software to be already installed on the victim's device. Also, we assigned the same transition probabilities to those hidden states, which had equal chances of occurring the next. For instance,  states $S2$  through $S6$ have equal probabilities to occur after the state S1. We ensured all the probabilities along each row in all of the matrices were assigned in such a way that the sum was equal to $1.0$. Figure \ref{fig:clickjacking} shows the state machine diagram for the attack type Tapjacking with initial, transition and emission probabilities where the hidden states are represented by yellow and observations in blue.

\subsection{Simulation Result}

Table \ref{tab:results} shows the results of best attack path sequence generated by Viterbi algorithm for observations $O$ taken from Table \ref{tab:obs_table} as an input. The idea behind giving known observation sequences was to check whether the algorithm is able to generate the corresponding attack path correctly or not.



\begin{table}[!ht]
{\footnotesize
\centering
\caption{Attack types identified based on the observation sequence input and the attack path generated by HMM.}
\label{tab:results}
\begin{tabular}{|p{2.7cm}|p{2.7cm}|c|} 
\hline
\multicolumn{1}{|c|}{\textbf{Observation Sequence}} & \multicolumn{1}{c|}{\textbf{Attack Path Generated}}     & \multicolumn{1}{c|}{\textbf{Attack Type}} \\ \hline
O1 $\rightarrow$ O2 $\rightarrow$ O3 $\rightarrow$ O4 & S1 $\rightarrow$ S2 $\rightarrow$ S10 $\rightarrow$ S11             & Clickjacking         \\ \hline
O1 $\rightarrow$ O5 $\rightarrow$ O6 $\rightarrow$ O4 & S1 $\rightarrow$ S6 $\rightarrow$ S10 $\rightarrow$ S11             & Tapjacking           \\ \hline
O7 $\rightarrow$ O8 $\rightarrow$ O9 $\rightarrow$ O1 $\rightarrow$ O10 $\rightarrow$ O4  & S1 $\rightarrow$ S3 $\rightarrow$ S7 $\rightarrow$ S9 $\rightarrow$ S10 $\rightarrow$ S11 & Scheme Squatting     \\ \hline
O7 $\rightarrow$ O11 $\rightarrow$ O9 $\rightarrow$ O1 $\rightarrow$ O10 $\rightarrow$ O4 & S1 $\rightarrow$ S5 $\rightarrow$ S8 $\rightarrow$ S9 $\rightarrow$ S10 $\rightarrow$ S11 & Task Impersonation   \\ \hline
O7 $\rightarrow$ O12 $\rightarrow$ O13 $\rightarrow$ O10 $\rightarrow$ O4 & S1 $\rightarrow$ S4 $\rightarrow$ S9 $\rightarrow$ S10 $\rightarrow$ S11       & Activity Hijack      \\ \hline
\end{tabular}
}
\end{table}

\begin{figure}[h]
 \centering
  \includegraphics[width=\linewidth]{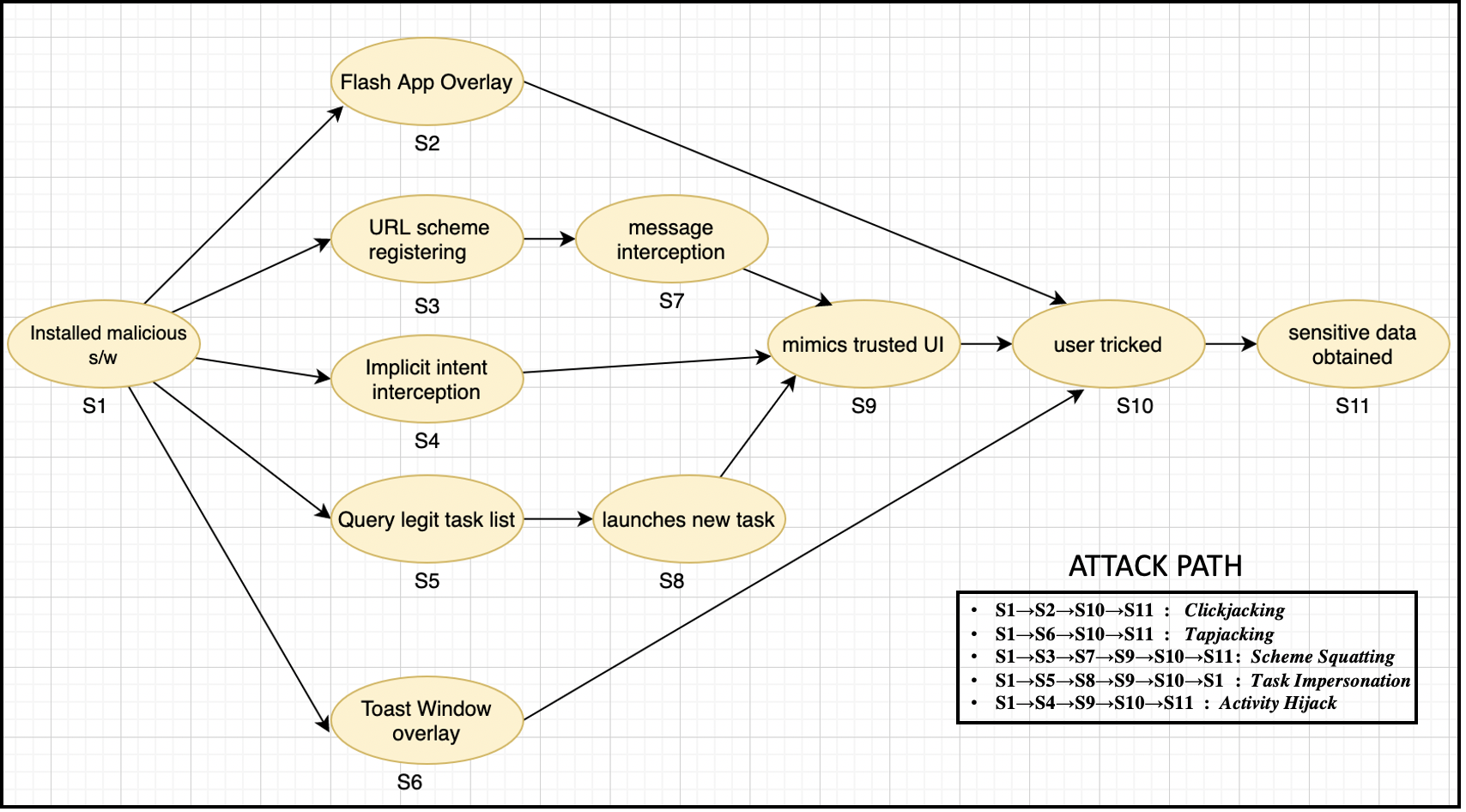}
  \caption{Attack paths and types generated by HMM.}
  \label{fig:HMM_model}
  \vspace*{-0.15in}
\end{figure}

Our HMM model generates an attack path (execution flow), given an observation sequence pertaining to an attack type using the Viterbi algorithm. We compared the path generated by the model against the execution flow descriptions described on the CAPEC's website for all the attack types. We identified the attack type that was closest to execution flow generated by the model. The corresponding state names to state number can be referred from Table \ref{tab:states_table}. Figure \ref{fig:HMM_model} shows the graphical representation of the various attacks paths, composed of hidden states of the model.

\section{Practical Implications}
\label{sec:practicalImplication}

In a realistic attack scenario, where a system is under attack, the observations can be captured by various command line utilities and software tools. 
In this section, we demonstrate the capturing process for some of the observations listed in Table \ref{tab:obs_table} using Windows and Android tools.

\begin{figure*}[!tbp]
  \centering
  \begin{minipage}[b]{0.60\textwidth}
    \includegraphics[width=\textwidth]{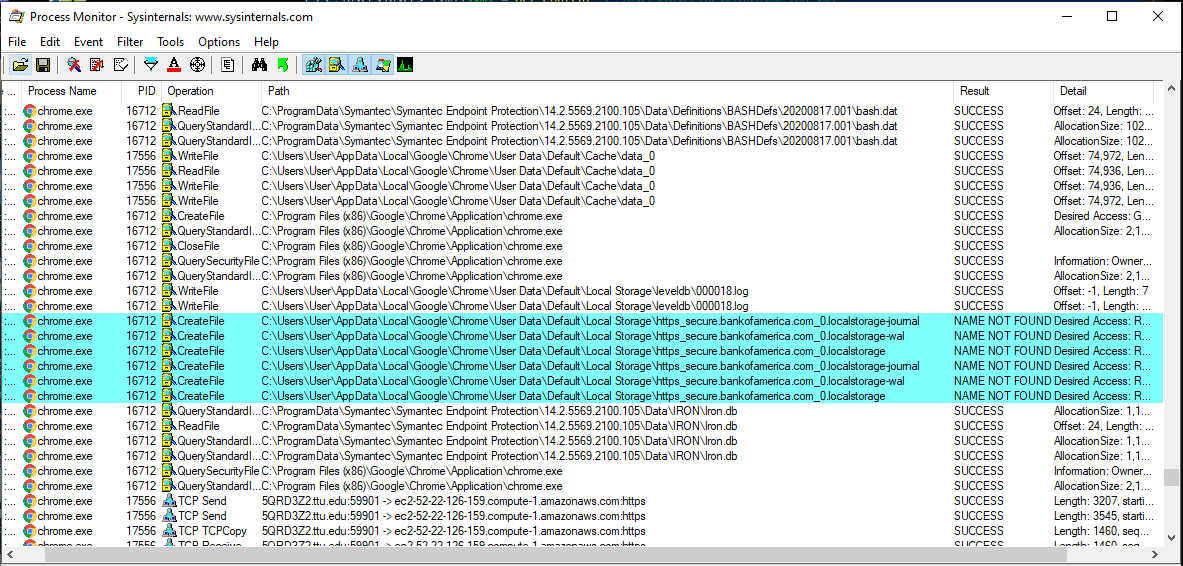}
    \caption{ProcMon User Interface.}
    \label{fig:GUI}
  \end{minipage}
  \hfill
    \begin{minipage}[b]{0.60\textwidth}
    \includegraphics[width=\textwidth]{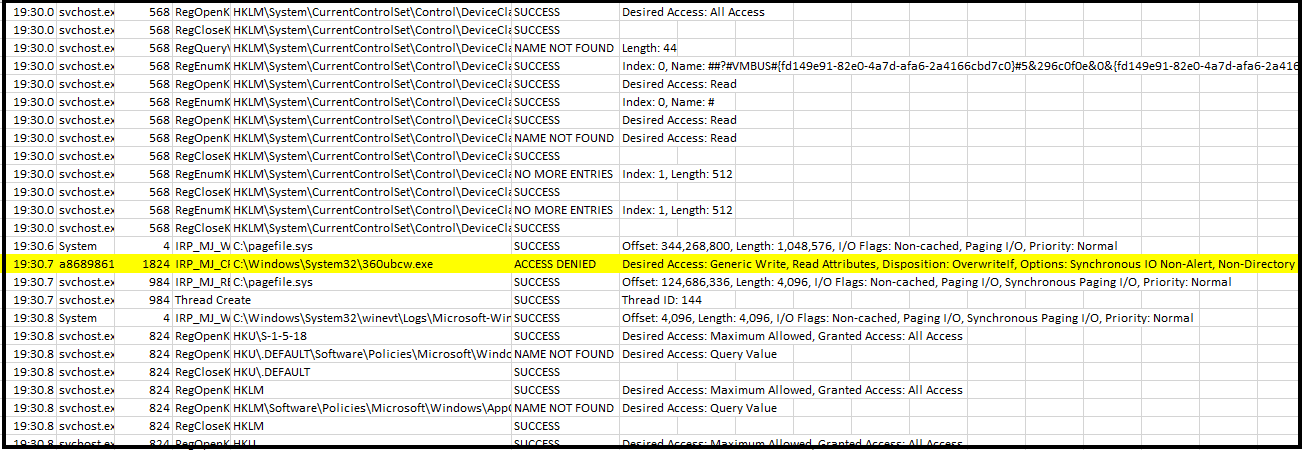}
    \caption{CSV log data with trace of file {\tt 360ubcw.exe} created by malware .}
    \label{fig:CSVlog}
  \end{minipage}
  \vspace*{-0.2in}
\end{figure*}

Process  Monitor (ProcMon) \cite{procmon} is a Windows-based utility that can capture real-time file system events, registry, and process/thread activity. The tool logs user activities performed on the local system as well as the interactions with the browser which includes process details, image path,  user and session ID. For instance, if the user opens the Chrome browser and visits the Bank of America's Website (i.e. observation O1 in Table \ref{tab:obs_table}) to do some transactions, the tool can capture these interactions and traces. Figure \ref{fig:GUI} shows the snapshots of user activity of visiting the bank website captured using ProcMon  (Highlighted in blue).

Any trace of malicious application running in the background (i.e. observation O7 in Table \ref{tab:obs_table}) can also be traced in the logs generated by the tool by applying ready-made filters of Process Monitor for Malware Analysis including \cite{eForensics}:
\begin{enumerate}
    \item {\it TCP/UDP Send and Receive,} Any connections that the malware may try to use while it is running.
    \item {\it Load Image,} DLL/Executable loading.
    \item {\it Create File,} new files being created.
    \item {\it Write/Delete/Rename File,} changes to the file system.
    \item {\it Registry Activities,} run entries for malware persistence.
\end{enumerate}

For instance, when a malicious spoofer\footnote{\scriptsize MD5: a86898615b642ed94adb1e361c30a8c06725b0cb396ab97b3ec08848c316a266} app runs in the background in Windows, it creates a file named {\tt 360ubcw.exe}, which is an indication of malware trying to plant a malicious file\footnote{https://virusshare.com/}. Figure \ref{fig:CSVlog} shows the screenshot of the file creation logs captured by ProcMon stored in a CSV file for this spoofer.

The instances of observation O2 in Table \ref{tab:obs_table}  can be captured by message tracking log command, \texttt{Get-MessageTrackingLog} in Powershell developed by Microsoft \cite{mtl}. To search the message tracking log entries for specific events, we can use the following syntax to capture the instances of the observations in the log file:

\vspace*{-0.1in}
{\small
\begin{verbatim}

Get-MessageTrackingLog 
    [-Server <ServerIdentity>] 
    [-ResultSize <Integer> | Unlimited] 
    [-Start <DateTime>] [-End <DateTime>] 
    [-EventId <EventId>] 
    [-InternalMessageId <InternalMessageId>] 
    [-MessageId <MessageId>] 
    [-MessageSubject <Subject>] 
    [-Recipients <RecipientAddress1, 
                  RecipientAddress2,...>] 
    [-Reference <Reference>] 
    [-Sender <SenderAddress>]
    
\end{verbatim}
}

To capture instances of observation O3 in Table \ref{tab:obs_table} and in order to detect an overlay which can occur both in a screen that appears on top of another application or window screen, a command line tool called \texttt{Monkey UI exerciser} \cite{Monkey} can be used to explore each app. The tool runs on device and generates UI event streams of user events such as clicks, touches, or gestures, as well as a number of system-level events . Whenever Monkey hits an overlay object, it records 
dynamic features listed 
\cite{yan2019understanding}. For launching the tool from command line, the {\tt adb} shell environment is required:
\begin{verbatim}
adb shell monkey [options] <event-count>    
\end{verbatim}

In a similar fashion, we can extract other observations using logs and other tools that can be used by the HMM model for attack prediction. It should be noted that we did not include the screenshots for rest of the observations described in Table \ref{tab:obs_table} for brevity.


\section{Conclusion and Future Work}
\label{sec:conclusion}

We presented a proof-of-concept of a prediction model based on Hidden Markov Model to identify the type of cyber attacks. The deriving idea is to capture sequences of observations and then map them out to the set of hidden states that form attack paths. All the probability matrices, i.e., transition, observation/emission and initial are manually determined based on the knowledge derived from the description of attacks from the CAPEC website.
The Viterbi algorithm implemented in the presented idea is capable of generating such attack paths and then matching them with the specification of a certain family of attacks. 

The current model deals with only one family of attacks where we have overlapping events representing the hidden states resulting in a smaller state search space.
We chose concrete attacks belonging to a single family of attacks because these concrete attacks share some common features. This makes modeling of attacks easy through HMM as compared to modeling concrete attacks which have no common features or characteristics. It  will be infeasible to model through a specific HMM model for each concrete attack. 

The model can be further expanded into an ensemble of HMM models trained on different attack families. Such an ensemble model can be scaled up to predict an extensive range of attacks. One of the limitations of this study is that it requires collection of vast number of observations belonging to different attacks using a variety of tools. This makes capturing such a vast pool of observations in real-time quite challenging. One of a possible solution to this problem could be to develop an integrated tool that can encapsulate observations from different user activities automatically. As a future work, our prediction model could help the victim trigger some defense actions against the attack if the model could predict the type of the ongoing attack in real time. We also demonstrated the feasibility of HMM through a case study. As future work, this model could be replicated on a larger set of attack types through collecting more data. The accuracy of the model can then be evaluated using metrics such as accuracy, precision, and recall.

\vspace*{-0.12in}
\section*{Acknowledgement}
This research work is supported by National Science Foundation (NSF) under Grant No. 1821560.

\bibliographystyle{IEEEtran}
\bibliography{IEEEabrv,mybibfile}

\begin{thebibliography}{10}
\providecommand{\url}[1]{#1}
\csname url@samestyle\endcsname
\providecommand{\newblock}{\relax}
\providecommand{\bibinfo}[2]{#2}
\providecommand{\BIBentrySTDinterwordspacing}{\spaceskip=0pt\relax}
\providecommand{\BIBentryALTinterwordstretchfactor}{4}
\providecommand{\BIBentryALTinterwordspacing}{\spaceskip=\fontdimen2\font plus
\BIBentryALTinterwordstretchfactor\fontdimen3\font minus
  \fontdimen4\font\relax}
\providecommand{\BIBforeignlanguage}[2]{{%
\expandafter\ifx\csname l@#1\endcsname\relax
\typeout{** WARNING: IEEEtran.bst: No hyphenation pattern has been}%
\typeout{** loaded for the language `#1'. Using the pattern for}%
\typeout{** the default language instead.}%
\else
\language=\csname l@#1\endcsname
\fi
#2}}
\providecommand{\BIBdecl}{\relax}
\BIBdecl

\bibitem{Holgado2020}
P.~{Holgado}, V.~A. {Villagrá}, and L.~{Vázquez}, ``Real-time multistep
  attack prediction based on hidden markov models,'' \emph{IEEE Transactions on
  Dependable and Secure Computing}, vol.~17, no.~1, Jan 2020.

\bibitem{Chen2012}
C.~{Chen}, D.~J. {Guan}, Y.~{Huang}, and Y.~{Ou}, ``Attack sequence detection
  in cloud using hidden markov model,'' in \emph{2012 Seventh Asia Joint
  Conference on Information Security}, Aug 2012, pp. 100--103.

\bibitem{be3899e815be45888cfc7ee529351b22}
A.~{Siami Namin} and S.~Dass, ``\BIBforeignlanguage{English}{A sensitivity
  analysis of evolutionary algorithms in generating secure configurations},''
  Dec. 2020.

\bibitem{9202661}
S.~Dass and A.~Siami~Namin, ``Evolutionary algorithms for vulnerability
  coverage,'' in \emph{2020 IEEE 44th Annual Computers, Software, and
  Applications Conference (COMPSAC)}, 2020, pp. 1795--1801.

\bibitem{Zan2009}
X.~{Zan}, F.~{Gao}, J.~{Han}, and Y.~{Sun}, ``A hidden markov model based
  framework for tracking and predicting of attack intention,'' in \emph{2009
  International Conference on Multimedia Information Networking and Security},
  Nov 2009, pp. 498--501.

\bibitem{Kholidy2014}
H.~A. {Kholidy}, A.~{Erradi}, S.~{Abdelwahed}, and A.~{Azab}, ``A finite state
  hidden markov model for predicting multistage attacks in cloud systems,'' in
  \emph{IEEE International Conference on Dependable, Autonomic and Secure
  Computing}, Aug 2014, pp. 14--19.

\bibitem{Ahmet2017}
A.~Okutan, S.~J. Yang, and K.~McConky, ``Predicting cyber attacks with bayesian
  networks using unconventional signals,'' in \emph{Proceedings of the 12th
  Annual Conference on Cyber and Information Security Research}, ser. CISRC
  '17, 2017.

\bibitem{Sanjana2020}
S.~Ingale, M.~Paraye, and D.~Ambawade, ``A survey on methodologies for
  multi-step attack prediction,'' in \emph{2020 Fourth International Conference
  on Inventive Systems and Control (ICISC)}, 2020, pp. 37--45.

\bibitem{Yu2007}
D.~Yu and D.~Frincke, ``Improving the quality of alerts and predicting
  intruder’s next goal with hidden colored petri-net,'' \emph{Computer
  Networks}, vol.~51, no.~3, pp. 632--654, 2007.

\bibitem{dong2020}
G.~Dong, J.~Wang, J.~Sun, Y.~Zhang, X.~Wang, T.~Dai, J.~S. Dong, and X.~Wang,
  ``Towards interpreting recurrent neural networks through probabilistic
  abstraction,'' 2020.

\bibitem{baum1966statistical}
L.~E. Baum and T.~Petrie, ``Statistical inference for probabilistic functions
  of finite state markov chains,'' \emph{The annals of mathematical
  statistics}, vol.~37, no.~6, pp. 1554--1563, 1966.

\bibitem{hmmtutorial}
L.~R. Rabiner, ``A tutorial on hidden markov models and selected applications
  in speech recognition,'' \emph{Proceedings of the IEEE}, vol.~77, no.~2, pp.
  257--286, Feb 1989.

\bibitem{capec}
MITRE, ``{CAPEC}-173: Action spoofing,'' Available at
  \url{https://capec.mitre.org/data/definitions/173.html}, 2019.

\bibitem{Chin2011}
E.~Chin, A.~P. Felt, K.~Greenwood, and D.~Wagner, ``Analyzing inter-application
  communication in android,'' in \emph{9th International Conference on Mobile
  Systems, Applications, and Services}, 2011, p. 239–252.

\bibitem{procmon}
\BIBentryALTinterwordspacing
M.~Russinovich. (2020, Sep) Process monitor v3.60. [Online]. Available:
  \url{https://docs.microsoft.com/en-us/sysinternals/downloads/procmon}
\BIBentrySTDinterwordspacing

\bibitem{eForensics}
eForensics Magazine, ``Dynamic malware analysis – process monitor and
  explorer,''
  https://eforensicsmag.com/dynamic-malware-analysis-process-monitor-and-explorer-by-prasanna-b-mundas/,
  2019.

\bibitem{mtl}
\BIBentryALTinterwordspacing
C.~Davis. Message tracking log. [Online]. Available:
  \url{https://docs.microsoft.com/en-us/exchange/mail-flow/transport-logs/search-message-tracking-logs?view=exchserver-2019}
\BIBentrySTDinterwordspacing

\bibitem{Monkey}
\BIBentryALTinterwordspacing
A.~Studio. {UI}/{A}pplication exerciser monkey. [Online]. Available:
  \url{https://developer.android.com/studio/test/monkey}
\BIBentrySTDinterwordspacing

\bibitem{yan2019understanding}
Y.~Yan, Z.~Li, Q.~A. Chen, C.~Wilson, T.~Xu, E.~Zhai, Y.~Li, and Y.~Liu,
  ``Understanding and detecting overlay-based android malware at market
  scales,'' in \emph{Annual International Conference on Mobile Systems,
  Applications, and Services}, 2019, pp. 168--179.

\end{thebibliography}
\end{document}